\documentclass[letterpaper]{article}
\usepackage[latin1]{inputenc}
\usepackage[american]{babel}
\usepackage{graphics}
\usepackage{amsmath}
\usepackage{times}

\makeatletter

 \newcommand{\lyxaddress}[1]{
   \par {\raggedright #1 
   \vspace{1.4em}
   \noindent\par}
 }

\makeatother

\newlength{\bekkoz}
\begin{document}

\title{Properties of a random attachment growing network}

\author{L\'{a}szl\'{o} Zal\'{a}nyi\( ^{1,2} \), G\'{a}bor Cs\'{a}rdi\( ^{1,2} \), Tam\'{a}s Kiss\( ^{1,2} \),
M\'{a}t\'{e} Lengyel\( ^{1,2} \),\\ Rebecca Warner\( ^{2,3} \), Jan
Tobochnik\( ^{2,3} \)\footnote{Corresponding author, e-mail: jant@kzoo.edu}\ \ and P\'{e}ter \'{E}rdi\( ^{1,2,3} \)}

\maketitle

\lyxaddress{\centering \( ^{1} \)\textit{Department of Biophysics KFKI Research
Institute for Particle and Nuclear Physics of the Hungarian Academy
of Sciences Budapest, Hungary}}

\lyxaddress{\centering \( ^{2} \)\textit{Center for Complex Systems Studies,
Kalamazoo College, Kalamazoo, MI 49006, USA}}

\lyxaddress{\centering \textit{\( ^{3} \)Physics Department, Kalamazoo College,
Kalamazoo, MI 49006, USA}}

\begin{abstract}
  In this study we introduce and analyze the statistical structural
  properties of a model of growing networks which may be relevant to
  social networks. At each step a new node is added which selects \( k
  \) possible partners from the existing network and joins them with
  probability \( \delta \) by undirected edges. The `activity' of the
  node ends here; it will get new partners only if it is selected by a
  newcomer. The model produces an infinite-order phase transition when
  a giant component appears at a specific value of \( \delta \), which
  depends on \( k \). The average component size is discontinuous at
  the transition. In contrast, the network behaves significantly
  different for \( k=1 \). There is no giant component formed for any
  \( \delta \) and thus in this sense there is no phase transition.
  However, the average component size diverges for \( \delta \geq
  \frac{1}{2} \).
\end{abstract}

\setlength{\parskip}{\bekkoz}

\section{Introduction}

There are many kinds of networks including probably the most
influential network of all, the World Wide Web \cite{baran64}. This
network is a popular one to analyze because of its size and easy
accessibility for statistical analysis. However, there are many other
networks that share some of the properties of the Web and some that do
not.  Among these networks we find social networks
\cite{milgram67,scott91,wasser94}, collaboration nets
\cite{watts98,watts99,newmann02,barabasi02}, industrial and business
related networks \cite{watts98,watts99,lux99}, transportation nets
\cite{banavar99} and many biological related nets such as food,
ecological, and protein interaction networks
\cite{jeong01,kauffman69,montoya00,sole00} and neural networks
\cite{sporns02}.

The mathematical description of networks started with the fundamental
works of Erd\H{o}s and R\'{e}nyi \cite{erdosr59,erdosr60}, which in the
absence of reliable data on large networks were rarely compared to
real networks. Recently, the computational boom has provided us an
increasing number of types of networks and more data on these networks.
One of the most exciting discoveries is the scale-free structures of
certain evolving networks \cite{faloutsos99,barabasi00,redner98}.  These
nets have power law degree distribution, where only a few vertices have
many connections to the others and the rest of the graph is rarely
connected. To explain the origin of this scale free structure of
networks Barab\'{a}si \textit{et al}.  \cite{barabasi99,barabasi992}
suggested the mechanism of preferential attachment and emphasized the
key role of growth. In their model the probability of a new node
connecting to an existing node is proportional to the degree of the
target node. Variations on this model include networks where there is
aging of nodes, nonlinear attachment probabilities, and re-wiring are
allowed. \cite{dorogovtsev00,dorogovtsev01,krapivsky01,krapivsky012}

Probably the most obvious feature of real networks that is missing
from most of the models studied by mathematicians and physicists are
characteristics of individual nodes in real networks which influence
the connection probability. Thus, if the nodes represent individual
persons, it is obvious that in many circumstances two people are more
likely to become connected in some form of relationship because of the
nature of their individual characteristics. Our model is motivated by
the need to incorporate this idea. A similar idea was used in a
preferential attachment model by Bianconi and Barab\'{a}si \cite{BB} who
assigned to each new node a fitness parameter. In their model a larger
fitness parameter may overcompensate the smaller probability of
attachment.

In our study we propose a simple model of growing networks whose
statistical properties are identical to a more complicated model
containing nodes with distinct characteristics. We will calculate the
edge distribution of the growing network, the distribution of cluster
sizes and the emergence of a giant cluster. We will also show how the
number of attempted connections made when a new node is added
determines the position and type of the phase transition as well as
the cluster size distribution.

\section{The Model}

We first consider a social network model where each node has
individual characteristics or traits. Each node that is added to the
network is assigned a permanent set of random traits which could be
coded as an ordered binary string or vector of length $L$. When a node
is added it chooses randomly \( k \in N\) possible partners from the
already existing nodes, or if there are less then \( k+1 \) (because
the simulation has not yet reached time step \( k+2 \)) it chooses all
the existing nodes as possible partners. A trait distance between the
new node and one of its possible partners is calculated based on their
trait vectors (\( \overrightarrow{t}_{1} \), \( \overrightarrow{t}_{2}
\)) using a distance measure, \( D\left( \overrightarrow{t}_{1},\,
  \overrightarrow{t}_{2}\right) \), such as the Hamming distance. Then
a connection is formed between the two nodes with a probability
determined from a given probability distribution over the distance
function \( p(D)\). Different functions, $p(D)$, correspond to
different soicopsychological situations. Thus, if we wish to model the
case where people are more likely to link together if they have
similar traits, then $p(D)$ would be a monotonically decreasing
function of $D$. For this case, the simplest $p(D)$ would be to form a
link if $D$ is below some threshold. This procedure is repeated for
each possible partner of the new node. Thus, each new node can have
initially up to $k$ links with the other existing nodes. Existing
nodes can have more than $k$ links as more nodes are added to the
network and link up with the existing nodes. There are no multiple
links between pairs of nodes.

\selectlanguage{american}

Because each node is given a random trait vector, and the nodes to
link to are also chosen randomly, many properties of the network
simply depend on the probability $\delta$, that two chosen nodes will
link together:

\begin{equation}
\label{deltadef}
\delta = \sum_{D} p\bigl(D(\overrightarrow{t}_{1},\,\overrightarrow{t}_{2})\bigr) r\bigl(D(\overrightarrow{t}_{1},\,\overrightarrow{t}_{2})\bigr)
\end{equation}

where $r(D)$ is the probability of the distance $D$ between two nodes, and the sum is over all possible
distance values. Thus, the model is reduced to the
following procedure. At each time step we add a node to the network,
and attempt to link with $k$ existing nodes which are chosen at
random. An actual connection is made with a probability $\delta$. The
asymptotic behavior of the network in the limit of large time \( t \),
does not depend on the initial condition of starting with a single
isolated vertex.

Although frequently structural properties of a network of nodes with trait
vectors depends only on $\delta$, there are other properties which
will depend on the detailed form of $p(D)$ and the nature of the trait
vectors. Examples of such properties include the distribution of
traits in different parts of the network and the correlation of traits
with distance in the network. For example, one can imagine a very
simple network of nodes representing men and women. In one network the
probability of forming a link is independent of sex, and in the other
persons prefer to link up with members of the opposite sex.  As long
as the mean probability of two chosen nodes linking together is the
same in the two scenarios the structural properties of the two
networks will be the same, but the distribution of men and women
within the network will be quite different in the two cases. In this
paper we confine ourselves to the structural properties of networks
and are considering these other non-structural properties in our
current research.

\section{Age dependence of the expected number of edges, and the edge distribution}

The expected number of edges at a node is approximately
\begin{equation}
\label{eq:kn}
K_{N}(t)=\overbrace{\delta k}^{\textrm{initial connections}}+
\overbrace{\sum _{s=N+1}^{t}\delta \frac{k}{s-1}}^{\textrm{later connections}}=\delta k\left(
1+H_{t-1}-H_{N-1}\right) \, ,
\end{equation}
where \( N>k \) is the time-step when it was created,
 (the smaller $N$ is, the older the node is)
\( t \) is the total simulation time,
\( \delta \) is the probability that two nodes form a connection, 
\(k \) is the maximum number of initial connections of a newly created
node, and \( H_{n} \) is the \( n^\textrm{th}\) harmonic number given by
the formula \( H_{n}=\sum ^{n}_{i=1}\frac{1}{i} \) for \( n>0 \), and \(
H_{0}=0 \). This equation shows that the number of edges of a node heavily
depends on the age of the node.

Equation~(\ref{eq:kn}) slightly overestimates the number of connections
for the oldest nodes in the network in two respects. First, the above
formula assumes that a node always has \( k \) possible initial
connections.  However, multiple connections between a pair of nodes
are not allowed, and there are less than \( k \) available partners
for the initial connections of a node created before or in the \( k
^\textrm{th} \) time step (overestimation of initial connections).
Second, the term for the late connections assumes that a node has a \(
k/(m-1) \) chance of being selected as the partner of the \( m
^\textrm{th} \) node (which chooses \( k \) possible partners out of
\( m-1 \) already existing nodes). However, for a node created in time
step \( N<k \), this term yields a probability of being chosen greater
than \( 1 \) between time steps \( N+1 \) and \( k \) (where \( m-1<k
\)) that is unacceptable again because multiple connections between a
pair of nodes are not allowed (overestimation of late connections). Below
is the formula correcting these errors, but will use the simpler,
uncorrected formula in the remaining part of our paper because the errors
are negligible.

\begin{equation}
\label{eq:kn_c}
\begin{array}{rl}
& K_{N}(t)= \delta \min \left( k,N-1\right) +\sum_{s=N+1}^{t}\delta \min \left(
\frac{k}{s-1},1\right) =\\
 & \\
& = \delta k\, \left( 1+H_{t-1}-H_{N-1}\right) -\\
 & -\delta \underbrace{\left( \underbrace{\max \left( k-N+1,0\right) }_{\textrm{initial connections}}+\underbrace{k\max \left( H_{k-1}-H_{N-1},0\right) -\max \left( k-N,0\right) }_{\textrm{late connections}}\right) }_{\textrm{correction for oldest nodes}}=\\
 & \\
 & \\
& =  \left\{ \begin{array}{ll}
\delta \, \left[ k\left( \alpha _{t}-\ln \left( k-1\right) -\frac{1}{2\left( K-1\right) }\right) -1\right]  & \textrm{if }N\leq k+1\\
\delta k\left( \alpha _{t}-\ln \left( N-1\right) -\frac{1}{2\left( N-1\right) }\right)  & \textrm{if }N>k+1\, ,
\end{array}\right. 
\end{array}
\end{equation}

using \( H_{n}\sim \ln n+\gamma +\frac{1}{2n} \),
where \( \gamma =-\int ^{\infty }_{0}e^{-x}\ln x\, dx\sim 0.5772 \) is
the Euler-Mascheroni constant \cite{conway96}, and \( \alpha
_{t}=1+\ln \left( t-1\right) +\frac{1}{2\left( t-1\right) } \).

\begin{figure}
{\centering\resizebox*{0.7\columnwidth}{!}{\includegraphics{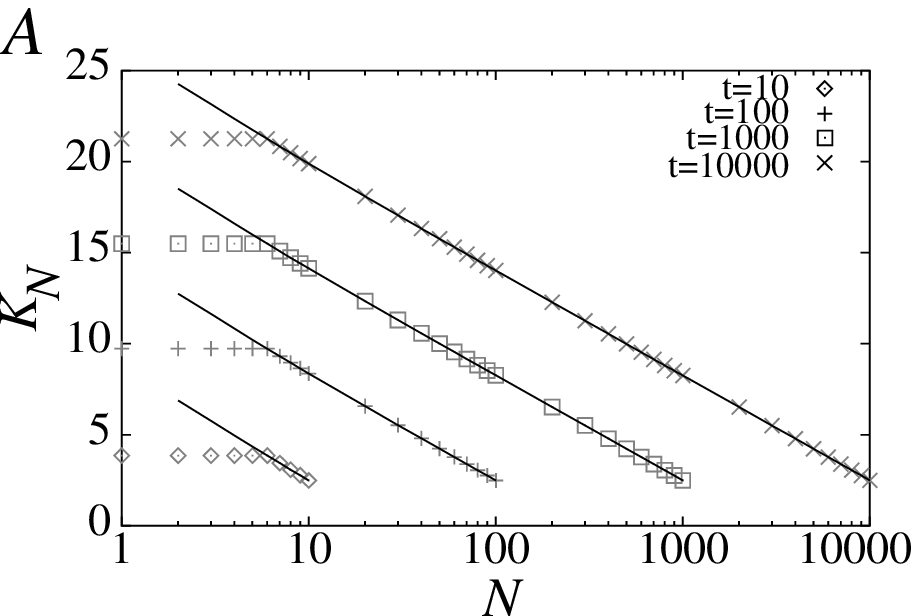}} \par}
{\centering\resizebox*{0.7\columnwidth}{!}{\includegraphics{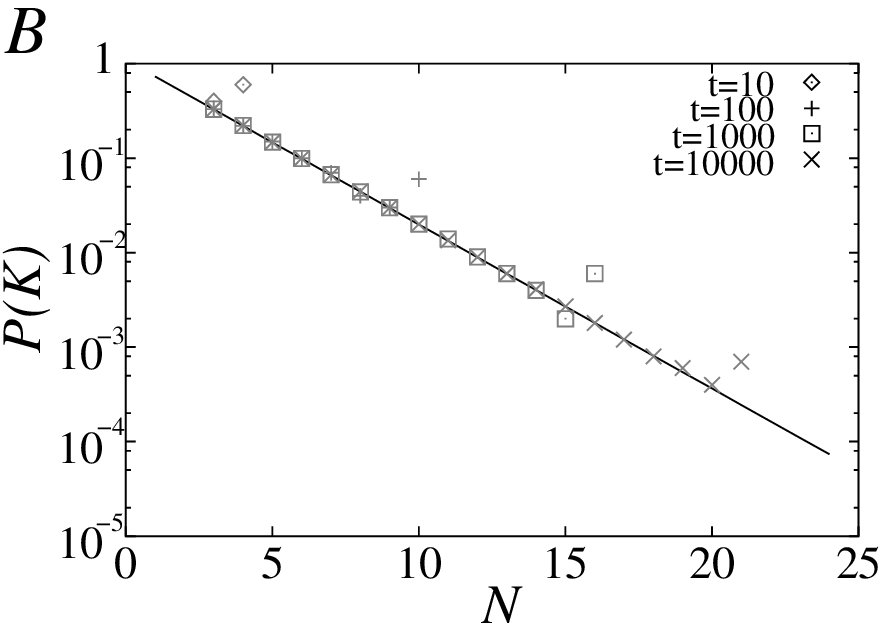}} \par}

\caption{\label{fig:kn_ca}\emph{(A)} Expected number of edges of a node
  (\protect\( K_{N}\protect \)) as a function of the age of the node
  (\protect\( N\protect \)) at different ages of the network
  (\protect\( t\protect \)).  \emph{Symbols} are numerically
  calculated values from Eq.\ (\ref{eq:kn_c}), showing that the first
  \protect\( k+1\protect \) nodes have the same number of connections
  at any \protect\( t\protect \), whereas there is an exponential
  break down in the expected number of edges for nodes created later
  than these.  \emph{Lines} represent approximations by Eq.\ 
  (\ref{eq:kn_a}): the number of edges of the first \protect\(
  k\protect \) nodes are overestimated because the correction term in
  Eq.\ (\ref{eq:kn_c}) was ignored (see text). Note that the x-axis is
  logarithmic. \emph{(B)} Edge distribution of the network.
  \emph{Symbols} represent numerically calculated distributions, where
  the numbers of edges of individual nodes were obtained from Eq.\ 
  (\ref{eq:kn_c}) These numbers were binned into integer values and the
  relative frequencies of occurrences in each bin were plotted. The
  \emph{line} represents the approximate distribution given by Eq.\ 
  (\ref{eq:pkn_a}), showing that it is valid for the edge distribution
  at large values of \protect\( t\protect \) (for \protect\( t\geq
  100\protect \)).  The mismatch between approximated and actual
  distributions at the highest connection numbers is due the same
  reason as in \emph{(A)}.  Note that y-axis is logarithmic.
  Parameters were \protect \( \delta =0.5 \protect \), \protect \( k=5
 \protect \). }
\end{figure}

\begin{figure}
{\centering\resizebox*{0.7\columnwidth}{!}{\includegraphics{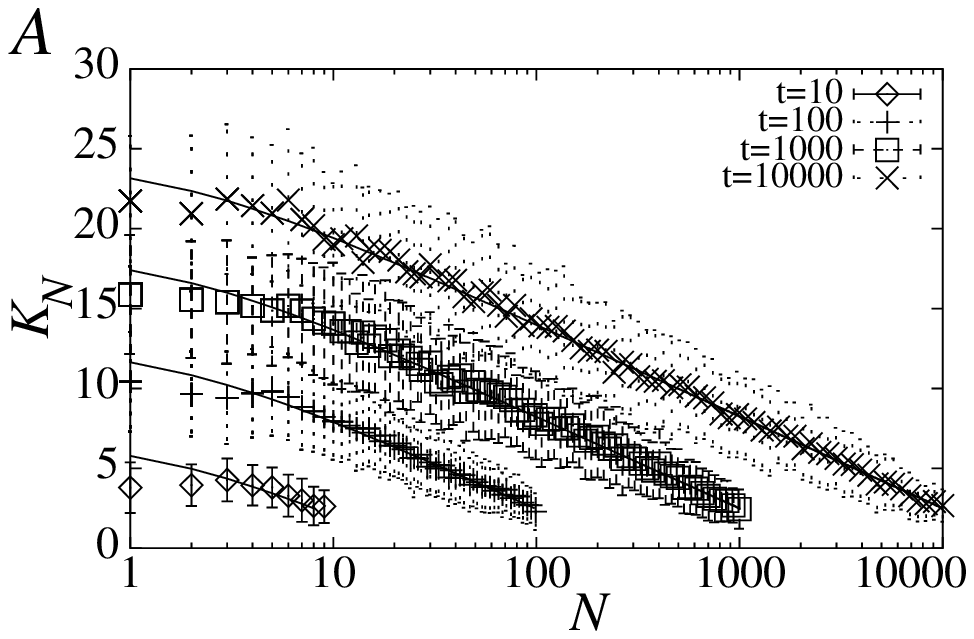}} \par}
{\centering\resizebox*{0.7\columnwidth}{!}{\includegraphics{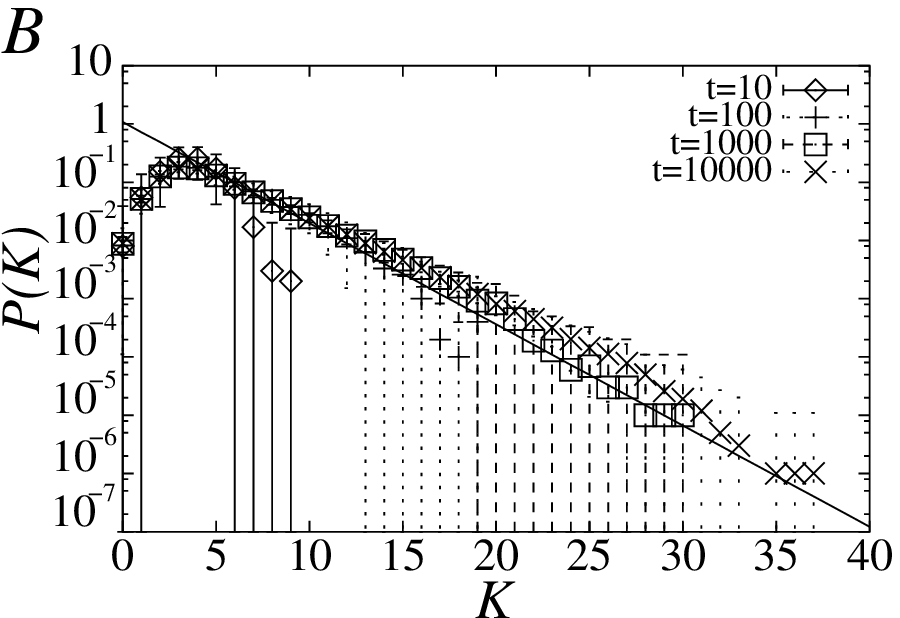}} \par}

\caption{\label{fig:kn_si}\emph{(A)} Numerical simulation of the average
  expected number of edges of a node (\protect\( K_{N}\protect \)) as
  a function of the age of the node (\protect\( N\protect \)).
  \emph{Symbols} are results of numerical simulations, \emph{line} is
  the graph of Eq.\ (\ref{eq:kn_a}). Also see notes of Fig.\ 
  \ref{fig:kn_ca}\emph{A}.  \emph{(B)} Numerical simulation of the
  edge distribution of the network (\emph{symbols}). \emph{Line} is
  the graph of Eq.~(\ref{eq:pkn_a}). Average relative
  frequency of individual number of edges and std.~ were calculated.
  Deviation of simulation from analytical results at high number of
  edges is a result of the finite size of simulated networks due to
  dispersion of expected number of edges arround its expected value as
  shown in part \emph{A} of this figure.  As age of the network
  increases this deviation disappears and simulation results approach
  analytical approximation for longer interval. At low number of edges
  deviation is a results from neglecting the correction term in
  Eq.~(\ref{eq:kn_a}) Averages and standard deviations were calculated
  from 100 simulations. Parameters were: \( \delta =0.5 \protect \),
  \protect \( k=5 \protect \) as in Fig.~\ref{fig:kn_ca}}

\end{figure}

Note that the first \( k+1 \) nodes are expected to have the same
number of connections (because \( K_{N} \) does not depend on \( N \)
in their case), and the edge number starts breaking down exponentially
for nodes created after time step \( k+1 \) (Fig.\ \ref{fig:kn_ca} and \ref{fig:kn_si}\emph{A}).
This means that this growth mechanism is identical to that where the first
\( k+1
\) nodes are created in the same time step.

We now wish to determine the edge distribution, $P(X)$, equal to the
probability that a node picked at random has on average $X$ edges. We
return to Eq.~(\ref{eq:kn}) ignoring the correction term in
Eq.~(\ref{eq:kn_c}), and write the formula for \( K_{N}(t) \) in the
simpler form (Fig.\ \ref{fig:kn_ca} and \ref{fig:kn_si}\emph{A}):
\begin{equation}
\label{eq:kn_a}
K_{N}\left( t\right) \simeq \delta k\left( \alpha _{t}-\ln \left( N-1\right) -\frac{1}{2\left( N-1\right) }\right) \, ,
\end{equation}
where \( \alpha _{t} \) is the same used in Eq.~(\ref{eq:kn_c}).
Using Eq.~(\ref{eq:kn_a}), neglecting the term $\frac{1}{2(N-1)}$ in
Eq.~(\ref{eq:kn_a}) for $N$ large enough, and knowing that the age
distribution of nodes is uniform, we analytically approximate the edge
distribution of the network with the following exponential
\begin{equation}
\label{eq:pkn_at}
P\left( K(t)=X\right) =\frac{1}{\delta kt}e^{-\frac{X}{\delta k}+\alpha _{t}}\, .
\end{equation}
We used the standard transformation rule for random variables, \(
P(N)=P(K_N)\left| \frac{dK_N}{dN}\right| \) with $P(N) = 1/t$. For
sufficiently large \( t \), due to the definition of \( \alpha _{t} \),
this can be effectively approximated by a
distribution which is independent of \( t \) (Fig.\ 
\ref{fig:kn_ca} and \ref{fig:kn_si}\emph{B}):
\begin{equation}
\label{eq:pkn_a}
P(X) =\frac{1}{\delta k}e^{-\frac{X}{\delta k}+1}\, .
\end{equation}

We can also determine a slightly different degree or edge distribution
which is the percentage of nodes with $m$ edges.  Denote by \(
d_{m}(t) \) the expected number of nodes with degree \( m \) at time
\( t \). The number of isolated nodes, \( d_{0}(t) \), will increase
by \( (1-\delta )^{k} \), which is the probability of the addition
node not connecting to any existing node, and decrease on average by
\( k\delta d_{0}(t)/t \) :
\begin{equation}
\label{eq:ed}
d_{0}(t+1)=d_{0}(t)+(1-\delta )^{k}-k\delta \frac{d_{0}(t)}{t}.
\end{equation}
The formula for the expected number of nodes of degree \( m>0 \) is a
bit complicated. For (\( 1\leq m\leq k \)) there are two ways to
increase \( d_{m} \): either selecting degree \( m-1 \) nodes for
connection with the new node or the new node having exactly \( m \)
edges. For (\( m > k \)), the new node cannot contribute to \( d_{m}
\). The decrease will be proportional to the probability of choosing a
degree \( m \) node for attachment.
\begin{eqnarray}
\nonumber 1\leq m\leq k:\quad  & d_{m}(t+1)= & \! d_{m}(t)+k\delta \frac{d_{m\! -\! 1}
(t)}{t}+\\
 & & +\left( \begin{array}{c}k\\
m
\end{array}\right) \delta ^{m}(1-\delta )^{k\! -\! m}-k\delta \frac{d_{m}(t)}{t}\label{eq:ed2} \\
m > k:\quad  & d_{m}(t+1)= & \! d_{m}(t)+k\delta \frac{d_{m\! -\! 1}(t)}{t}-k\delta
\frac{d_{m}(t)}{t}.\label{eq:ed3} 
\end{eqnarray}

These equations are correct as \( t\rightarrow \infty \), and
numerical simulations show that \( d_{m}(t)\sim p_{m}t \).
Substituting this form into the equations for $d_m(t)$ we
obtain

\begin{eqnarray} m\leq k:\quad & p_{m}= & \delta ^{m}\sum
  ^{m}_{j\! =\! 0}\left( \begin{array}{c}
      k\\
      j
\end{array}\right) \frac{(1-\delta )^{k\! -\! j}}{(1+k\delta )}\left( \frac{k}{1+k\delta }\right) ^{m\! -\! j},\label{eq:ed4} \\
m > k:\quad  & p_{m}= & p_{k}\left( \frac{k\delta }{1+k\delta }\right) ^{m\! -\! k}.\label{eq:ed5} 
\end{eqnarray}

This degree distribution \( p_{m} \) decays exponentially consistent
with our previous result for $P(X)$.

\section{Critical behavior}

\subsection{\label{sec:cluster}Cluster size distribution}

In some network models, such as the preferential attachment models,
all the nodes belong to a single cluster. For such models the focus is
on the degree distribution and the distance between nodes in the
network. However, our network can contain a number of disconnected
clusters of nodes. Then the key questions become what is the cluster
size distribution and is there a phase transition between a collection
of finite size clusters and the appearance of a giant cluster much
larger than the rest.  The transition is similar to that in
percolation, with our parameter $\delta$ playing the role of the site
occupation probability in a percolation model. The key difference
between our model and percolation models is that our nodes do not sit
on a lattice structure, and there is thus no geometric constraints.
The definition of a giant cluster in our model is somewhat different
than a spanning cluster in percolation models. Nevertheless, some of
the behavior is similar.

Our model is similar to one by Calloway \textit{et al}.
\cite{callaway01} where an infinite order phase transition was found.
In that model after a node was added to the network, two nodes were
picked at random and connected with probability $\delta$. Our model is
more general in that we consider the effect of making more than one
link at any given time.  Also, in our model the new links are between the
added node and exisiting nodes, whereas in the model by Calloway
\textit{et al} the new links are between any two nodes in the
network. 

To determine the cluster distribution we use a procedure similar to
the one we used to calculate the degree distribution. The cluster
number \( N_{j}(t) \) denotes the expected number of clusters of size
$j$. On average, at each time step, \( (1-\delta )^{k} \) isolated
nodes arrive at the network and \( k\delta N_{1}(t)/t \) nodes will be
chosen for attachment reducing $N_1$. Thus, \( N_{1} \) is described
by
\begin{equation}
\label{eq:cd}
N_{1}(t+1)=N_{1}(t)+(1-\delta )^{k}-k\delta \frac{N_{1}(t)}{t}.
\end{equation}
For \( j>1 \) new clusters of size $j$ come from connecting the new
node to a cluster of size $j-1$ or if $k > 1$ using the new node to
make connections between smaller clusters whose sizes add up to $j$.
Reducing $N_j$ will be \( jk\delta N_{j}(t)/t \) nodes from clusters
of size $j$ connecting to the new node. Thus, we have
\begin{equation}
\label{eq:cd2}
N_{2}(t+1)=N_{2}(t)+\left( \begin{array}{c}
k\\
1
\end{array}\right) \delta (1-\delta )^{k\! -\! 1}\frac{N_{1}(t)}{t}-k\delta \frac{2N_{2}(t)}{t}
\end{equation}
\[
\vdots \]
\[
N_{j}(t+1)=N_{j}(t)+ \left( \sum _{r\! =\! 1}^{\min (k,j-1)}\left( \begin{array}{c}
      k\\
      r
  \end{array}\right) \delta ^{r}(1-\delta )^{k\! -\! r}\times \right.\]
\[  \left. \times \sum_{\substack{z_1+\ldots+z_r=j-1\\ z_i\ge 1,\,\,i\le r}}
\frac{z_{1}N_{z_{1}}(t)}{t}\frac{z_{2}N_{z_{2}}(t)}{t}\cdots
\frac{\left( j-1-\sum ^{r-1}_{i=1}z_{i}\right) N_{\left( j-1-\sum
      ^{r-1}_{i=1}z_{i}\right) }(t)}{t}\right) - \]


\begin{equation}
  \label{eq:cd3}
  -k\delta \frac{jN_{j}(t)}{t}.
\end{equation}

\begin{figure}
\resizebox{\textwidth}{!}{\includegraphics{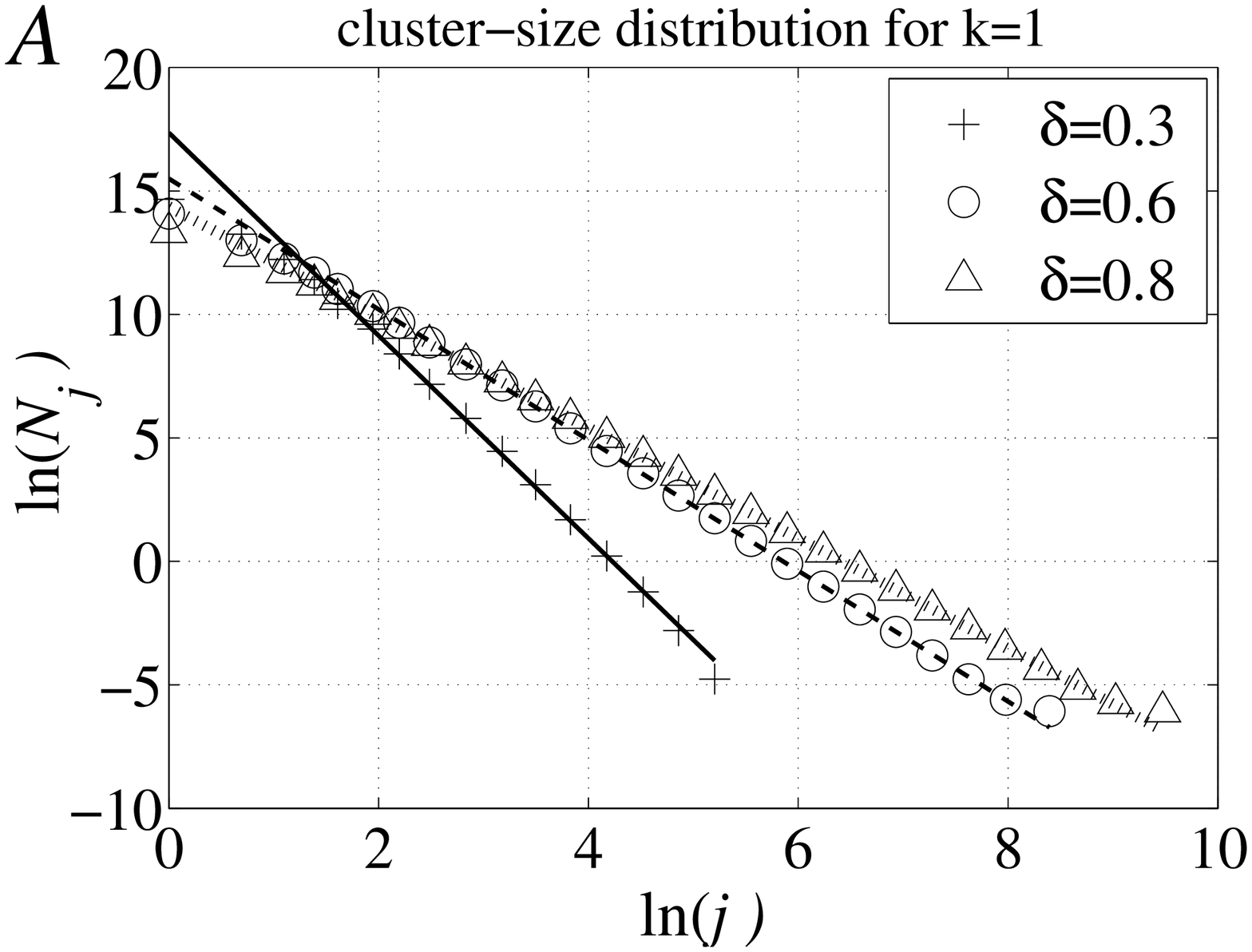} \includegraphics{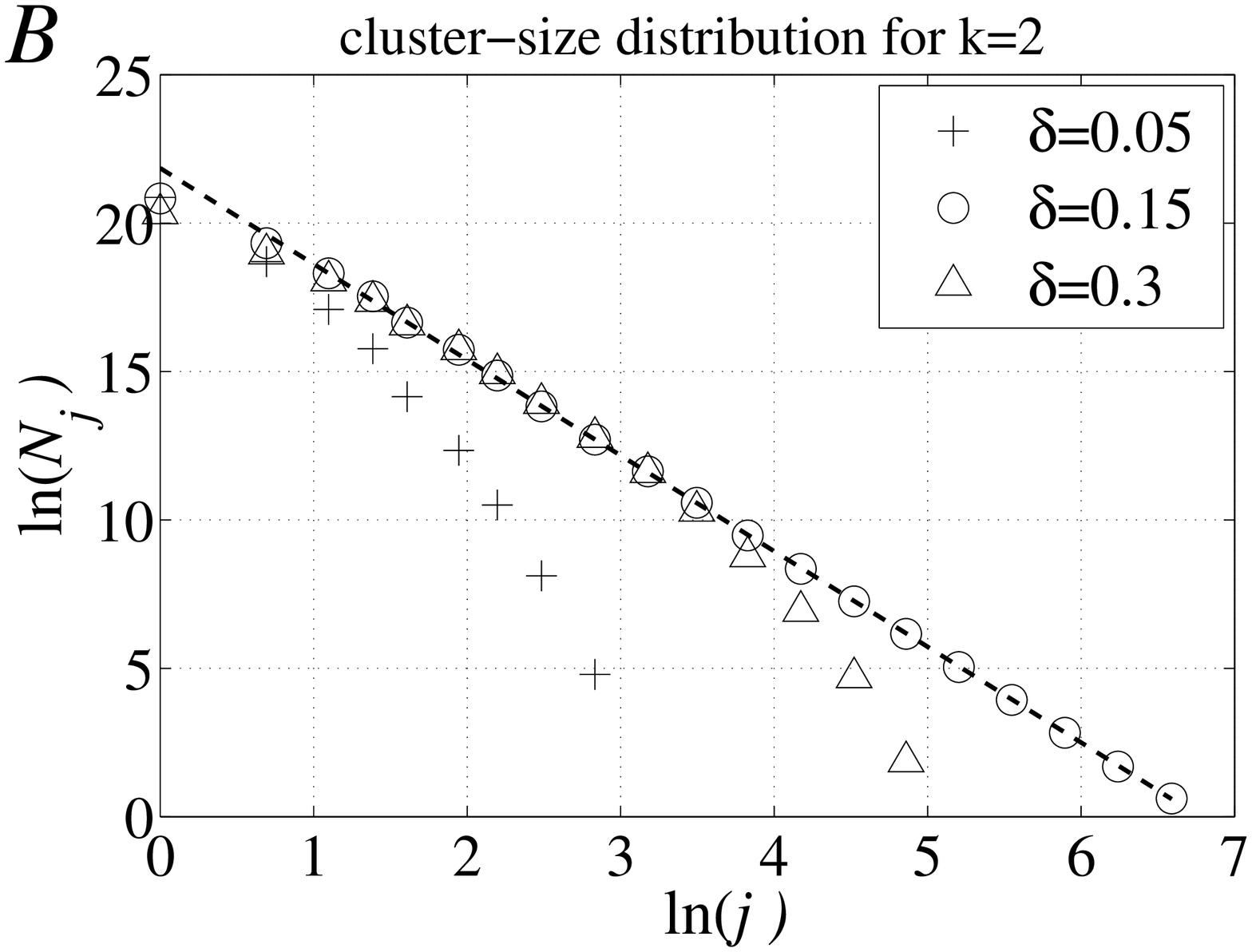}}
\caption{\label{fig:clx} Cluster size distribution for different \protect\( \delta \protect \)-s
  and \protect\( k=1\protect \) (left), \protect\( k=2\protect \)
  (right). Solid, dashed and dotted lines are obtained from a least
  squares fit for the interval \protect\( 11>\ln (N_{j})>-4\protect \)
  \emph{(A)} and \protect\( 20>\ln (N_{j})>2\protect \) \emph{(B)} indicating
  the power-law behavior of the distributions. Simulation data were
  obtained by averaging over 500 runs of \protect\( 10^{7}\protect \)
  time-steps and are shown on a log-log plot. Note that in figure
  \emph{(B)} simulations for \protect\( \delta = 0.05 \protect \) and
  \protect\( \delta = 0.3 \protect \) distributions do not follow a
  power-law. In Section \ref{sec:phase} it is shown that there is a
  phase transition near \protect\( \delta =0.146\protect \).}
\end{figure}

The first sum in Eq.\ (\ref{eq:cd3}) determines the number of sums in the
next term. Each of these sums represent a cluster that is melted
into the $j$ sized cluster. These equations are valid for \(
t\rightarrow \infty \), where the probability of closed loops tends to
zero. The giant cluster, if there exists one, is an exception in which
connection of nodes in loops is not negligible.  Thus,
Eq.~(\ref{eq:cd3}) holds only for the finite sized clusters in the
network. This property lets us determine a generating function which
we can use to find the size of the giant cluster.  Our simulations
show that solutions of Eqs.~(\ref{eq:cd}), (\ref{eq:cd2}) and
(\ref{eq:cd3}) are of the steady state form \( N_{j}(t)=a_{j}t \).
Using this form in Eqs.~(\ref{eq:cd}), (\ref{eq:cd2}) and
(\ref{eq:cd3}), we find
\begin{eqnarray}
a_{1} & = & \frac{(1-\delta )^{k}}{1+k\delta }\label{eq:cd4} \\
a_{2} & = & \frac{\left( \begin{array}{c}
k\\
1
\end{array}\right) \delta (1-\delta )^{k\! -\! 1}a_{1}}{(1+2k\delta )}\label{eq:cd5} 
\end{eqnarray}
\[a_{j}=\frac{1}{1+jk\delta }\left( \sum _{r\! =\! 1}^{\min (k,j-1)}\left( \begin{array}{c}
      k\\
      r
    \end{array}\right) \delta ^{r}(1-\delta )^{k\! -\! r}\times \right.\]
\begin{equation}
  \label{eq:cd6}
  \left. \times \sum_{\substack{z_1+\ldots+z_r=j-1\\ z_i\ge 1,\,\,i\le r}}
     \left( j-1-\sum ^{r-1}_{i=1}z_{i}\right) a_{\left( j-1-\sum ^{r-1}_{i=1}z_{i}\right) } \prod_{l=1}^{r-1}\left(z_l a_{z_l}\right) \right) .
\end{equation}
Generally we cannot obtain a simpler equation for the cluster size
distribution \( a_{j} \), except for \( k=1 \). Substituting \( k=1 \)
into the Eqs.~(\ref{eq:cd4}), (\ref{eq:cd5}) and (\ref{eq:cd6}) we
obtain after some algebra the general result
\begin{equation}
\label{eq:cd1k1}
a_{j}=(1-\delta )\delta ^{j\! -\! 1}(j-1)!\prod _{m=1}^{j}\frac{1}{1+m\delta },
\end{equation}
which can be written in the form:
\begin{equation}
\label{eq:cd2k1}
a_{j}=\frac{(1-\delta )\Gamma (1\! /\! \delta )}{\delta ^{2}}\frac{\Gamma (j)}{\Gamma (j+1+1\! /\! \delta )},
\end{equation}
where \( \Gamma (x) \) denotes the gamma-function.
Eq.~(\ref{eq:cd2k1}) shows that the cluster size distribution for
\( k=1 \) always follows a power-law distribution. This result is
confirmed by simulations shown in the left graph of Fig.\ 
\ref{fig:clx}. Distributions of cluster sizes for \( k=2 \) (right
graph of Fig.\ \ref{fig:clx}), in contrast to \( k=1 \) show power-law
behavior only near the phase transition.

\setlength{\parskip}{\bekkoz}

\subsection{\label{sec:phase}Position of the phase transition}

\begin{figure}
{\centering\resizebox*{0.7\textwidth}{0.4\textheight}{\includegraphics{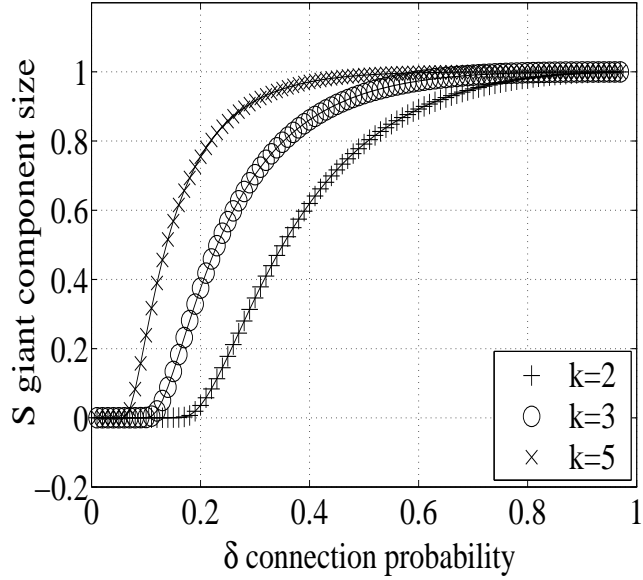}}\par}
\caption{\label{fig:S1}Giant cluster size \protect\( S\protect \) as a
  function of \protect\( \delta \protect \) and \protect\( k\protect
  \).  Symbols are from simulations of the growing network for
  \protect\( 10^{6}\protect \) time steps averaged over 30 runs. Lines
  are from the analytical calculations.}
\end{figure}

Fig.\ \ref{fig:S1} shows the simulation results for \( S \), the ratio
of the average size of the largest cluster to the total number of
nodes versus the connection probability \( \delta \). The figure
suggests that there is a smooth transition in the appearance of \( S \)
at a specific value of $\delta$ between \( \delta =0 \) and \( \delta
=0.2, \) which depends on the parameter \( k. \) To predict the
position of a possible phase transition \( \delta _{c} \)
\cite{callaway01}, we will use a generating function for the cluster
size distribution \cite{wilf94}. To derive the generating function we
use the iterative Eqs.~(\ref{eq:cd4}), (\ref{eq:cd5}), and
(\ref{eq:cd6}).  The generating function will be of the form:
\begin{equation}
\label{eq:gf1}
g(x)=\sum ^{\infty }_{j=1}b_{j}x^{j},
\end{equation}
where \begin{equation}
\label{eq:gf2}
b_{j}=ja_{j},
\end{equation}
is the probability that a randomly chosen node is from a cluster of
size \( j \). Multiplying both size of Eqs.~(\ref{eq:cd4}),
(\ref{eq:cd5}) and (\ref{eq:cd6}) by \( jx^{j} \), and summing over \(
j \) we derive a differential equation for \( g(x) \)
\begin{equation}
\label{eq:gf3}
g=-k\delta g\prime +x(1-\delta )^{k}+\sum ^{k}_{i=1}\left( \begin{array}{c}
k\\
i
\end{array}\right) \delta ^{i}(1-\delta )^{k\! -\! i}(x^{2}g\prime g^{i\! -\! 1}i+xg^{i}).
\end{equation}
Rearranging for \( g\prime \) we obtain
\begin{equation}
\label{eq:gf4}
g\prime =\frac{(1-\delta )^{k}-g/x+\sum ^{k}_{i=1}\left( \begin{array}{c}
k\\
i
\end{array}\right) \delta ^{i}(1-\delta )^{k\! -\! i}g^{i}}{k\delta -x\sum ^{k}_{i=1}\left( \begin{array}{c}
k\\
i
\end{array}\right) \delta ^{i}(1-\delta )^{k\! -\! i}g^{i\! -\! 1}i},
\end{equation}
which can be further simplified to
\begin{equation}
\label{eq:gf5}
g\prime =\frac{-g/x+(1+(g-1)\delta )^{k}}{k\delta -xk\delta (1+(g-1)\delta )^{k\! -\! 1}}.
\end{equation}

The generating function for the finite size clusters is exactly one at
\( x=1 \) when there is no giant cluster in the network and \( g(1)<1
\) otherwise. Hence
\begin{equation}
\label{eq:gf6}
S=1-g(1).
\end{equation}
Without an analytic solution for Eqs.~(\ref{eq:gf5}), we calculate \(
S \) numerically by integrating Eqs.~(\ref{eq:gf5}) with the initial
condition \( (x,g(x))=(x_{0},\, x_{0}(1-\delta )^{k}/(1+k\delta )) \)
where \( x_{0} \) is small. This is equivalent to starting with a
cluster of only one node.  In Fig.\ \ref{fig:S1} there are results from
direct simulations of the model (symbols) and solid lines from the
integration of the generating function. The agreement is good which
verifies the approximations.

\begin{figure}
{\centering\resizebox*{1\textwidth}{0.4\textheight}{\includegraphics{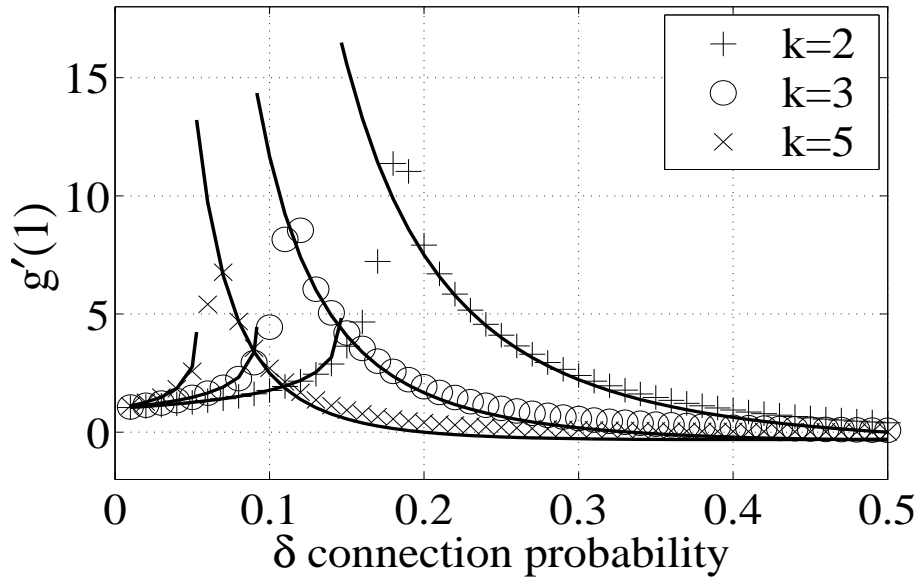}}\par}

\caption{\label{fig:g1}Discontinuity in \protect\( g\prime (1)\protect \)
  for different values of \protect\( k\protect \). Solid lines are
  theoretical, and symbols are results from the simulations of growing
  networks for \protect\( 10^{6}\protect \) time steps, averaged over
  30 runs.}
\end{figure}

To discuss the phase transition location we first consider the cases
\( k>1 \). Consider the expected value that a randomly chosen node
belongs to a finite size cluster. We can determine this quantity in
terms of the generating function \( g(x) \)
\begin{equation}
\label{eq:pt1}
\left\langle s\right\rangle =\frac{g\prime (1)}{g(1)}.
\end{equation}
For those values of \( \delta \) where no giant cluster exists, \(
\delta <\delta _{c} \), \( g(1)=1, \) and both the numerator and
denominator of Eq.~(\ref{eq:gf5}) goes to zero as \( x\rightarrow 1
\).  Using L'Hopital's rule we derive a quadratic equation for \(
g\prime (1) \).  The solution of this equation is
\begin{equation}
\label{eq:pt2}
g\prime (1)=\frac{1-2k\delta \pm \sqrt{(2k\delta -1)^{2}-4k(k-1)\delta ^{2}}}{2k(k-1)\delta ^{2}},
\end{equation}
for \( g(1)=1 \).  Because as \( \delta \rightarrow 0 \) all
clusters will have size 1, one can show that the correct solution of
Eq.~(\ref{eq:pt2}) is the one with the negative sign. In addition from
Eq.~(\ref{eq:pt2}) we can find the location of the phase transition. It is
the value of \( \delta \) where the solution of Eq.\ref{eq:pt2} becomes
complex:
\begin{equation}
\label{eq:pt3}
\delta _{c}=\frac{1-\sqrt{1-1/k}}{2}.
\end{equation}
In the region where there is a giant cluster \( \delta >\delta _{c}
\), Eq.~(\ref{eq:gf5}) becomes as \( x\rightarrow 1, \)
\begin{equation}
\label{eq:pt4}
g\prime =\frac{-g+(1+(g-1)\delta )^{k}}{k\delta -k\delta (1+(g-1)\delta )^{k\! -\! 1}},
\end{equation}
which is still not solvable analytically. Making the approximation \(
(1\pm a)^{k}\approx 1\pm ka \) when \( a\ll 1 \) , we can simplify
Eq.~(\ref{eq:pt4}) close to \( \delta _{c} \):
\begin{equation}
\label{eq:pt5}
g\prime (1)\approx \frac{k\delta -1}{k\delta ^{2}(1-k)},
\end{equation}
where \( g(1)<1 \), \( \delta >\delta _{c} \), and \( (g(1)-1)\delta
\ll 1 \). In Fig.\ \ref{fig:g1} we show the simulation results and the
above derived theoretical functions for \( g\prime (1) \).  We can see
that for \( \delta <\delta _{c} \), where we have an explicit
expression for \( g\prime (1) \) in terms of the parameters \( k \)
and \( \delta \) the fit is very good. For \( \delta >\delta _{c} \)
the fit is good close to the phase transition point, where the
approximation \( (g-1)\delta \ll 1 \) holds. Although below \( \delta
_{c} \) the description of \( g\prime (1) \) is very good, it seems
that the location of the phase transition and the value of the
function \( g\prime (1) \) above \( \delta _{c} \) is somewhat
different than the data. Also if we carefully check Fig.\ \ref{fig:g1}
at the jumps, we find that the larger the jump the less accurate the
theory seems to be. This can be explained as follows.  At the critical
point the average size of finite clusters jumps, hence much larger
clusters appear in the network. As we can only simulate for a finite
time large (but not the giant) clusters are underrepresented. The
weights of them computed from the simulation data are less then they
would be in an infinitely long simulation.  Away from the transition
regime fewer finite size clusters remain beside the giant cluster in
the network, and thus the distribution can be specified better.

Although the formalism using the generating function can be done for
\( k=1 \), the meaning of a giant cluster is problematic. In Section
\ref{sec:cluster} we showed that the size-distribution of clusters for
\( k=1 \) always follows a power-law which means there is no obvious
border between the `giant' cluster and smaller clusters.  There is not
a sharp break between the largest and the next largest cluster. The
physical reason for this is that clusters grow only by the addition of
newly added nodes. This is different than the case for $k> 1$ and in
percolation models where clusters can also grow by a link combining two
clusters.  In this sense no giant cluster appears in the network except
for
\(
\delta =1 \).  Eq.~(\ref{eq:gf5}) becomes
\begin{equation}
\label{eq:pt6}
g\prime (x)=\frac{(1-\delta )-g/x+\delta g}{\delta (1-x)},
\end{equation}
which becomes \( \frac{0}{0} \) in the limit \( x\rightarrow 1 \) with
\( g(1)=1 \). Applying L'Hopital's rule yields
\begin{equation}
\label{eq:pt7}
g\prime (1)=\frac{1}{1-2\delta }.
\end{equation}
At \( \delta =\frac{1}{2} \), \( g\prime (1)\rightarrow \)\( \infty
\), which means the average size of finite clusters approaches
infinity. From the definition of \( g(x) \) in Eq.~(\ref{eq:gf1}) and
the power-law cluster size distribution for \( a_{j} \), it follows
that \( g(1)=1 \) for any \( \delta \neq 1 \). To see that \( g\prime
(1) \)\( \rightarrow \infty \) as \( x\rightarrow 1 \) for \( \delta
>\frac{1}{2} \), we consider the sum form of the generating function
in Eq.~(\ref{eq:gf1}).  For large \( j \), \( a_{j}\approx
\frac{1}{j^{(1+1\! /\! \delta )}} \) Eq.~(\ref{eq:cd2k1}), and \(
g\prime =\sum _{j\! =\! 1}^{\infty }j^{2}a_{j}x^{j-1} \), which can
not be summed for \( \delta \geq \frac{1}{2}. \)

When \( \delta <\frac{1}{2}, \) the probability of a new node not
joining a cluster is higher then joining, and thus the weight of small
clusters is higher than that of larger clusters, and hence the average
size remains finite. As \( \delta \rightarrow \frac{1}{2} \), the
probability of forming clusters increases and so do the weight of
large clusters.

\subsection{Infinite-order transition}

\begin{figure}
{\centering\resizebox*{0.8\textwidth}{0.4\textheight}{\includegraphics{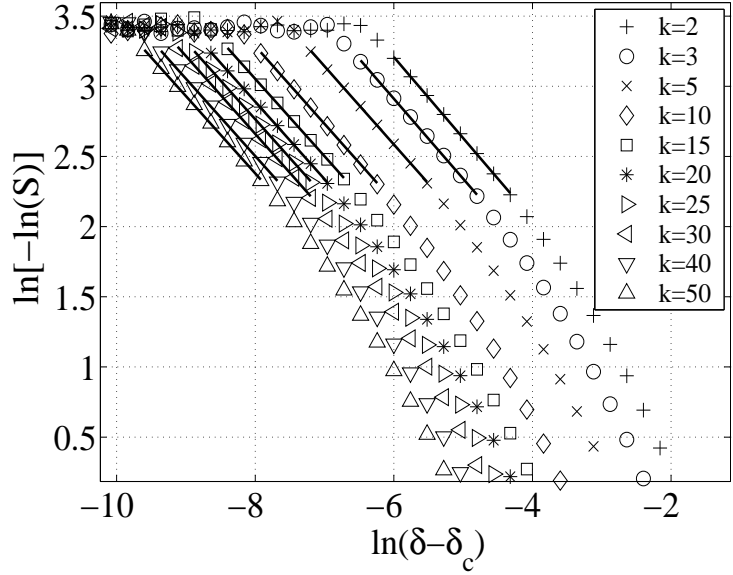}}\par}

\caption{\label{fig:loglogS} Numerical calculation of the giant cluster size
  close to but above the phase transition. Least-squares fitted solid
  straight lines suggest \protect\( S(\delta )\sim e^{\alpha (\delta
    -\delta _{c})^{\beta }}\protect \).  The flat ends of the curves
  on the top appear due to the limit of the accuracy of numerical
  integration.}
\end{figure}

To show the nature of our phase transitions \cite{callaway01}, we
numerically integrated Eq.~(\ref{eq:gf5}) for different values of \( k \)
near the corresponding critical \( \delta _{c} \). In Fig.\
\ref{fig:loglogS} the linear parts of the log(-log(S)) plots suggest that
\begin{equation}
\label{eq:io1}
S(\delta )\sim e^{\alpha (\delta -\delta _{c})^{\beta }}\qquad \textrm{as }\delta \rightarrow \delta _{c},
\end{equation}
and because all derivatives of \( S \) vanish at \( \delta _{c} \), the
transition is of infinite order.

Table \ref{tbl:fl} contains the parameters of the fitted straight
lines in Fig.\ \ref{fig:loglogS}. As the calculations were done close
to the numerical limit and referring to the similar results in
\cite{callaway01} we conjecture that \( \beta \) equals \(
-\frac{1}{2} \) for all \( k \). This result suggests that the
mechanism of the transition is common and the number of possible
partners for each node to link to determines the speed of emergence of
the giant cluster \( S. \) These results are in accord with
Eq.~(\ref{eq:pt5}), the average cluster size decrease is approximately
independent of \( k, \) but the size of the jump and the rate of
decrease is driven by \( k \).

\begin{table}
\scriptsize
{\centering \begin{tabular}{|c||c|c|c|c|c|c|c|c|c|c|}
\hline 
\( k \)&
2&
3&
5&
10&
15&
20&
25&
30&
40&
50\\
\hline 
\( \alpha  \)&
-0.25&
-0.5&
-0.75&
-1.14&
-1.35&
-1.52&
-1.64&
-1.77&
-1.9&
-2.02\\
\hline 
\( \beta  \)&
-0.577&
-0.569&
-0.557&
-0.554&
-0.551&
-0.552&
-0.551&
-0.554&
-0.551&
-0.55\\
\hline
\end{tabular}\par}

\caption{\label{tbl:fl}The parameter values (\protect\( \alpha \textrm{ and }\beta \protect \))
of the fitted lines in Fig.\ \ref{fig:loglogS}. Taking into account
that we were at the border of the maximal numerical accuracy and that
the fit is short we presume \protect\( \beta =-\frac{1}{2}\protect \).}
\end{table}

\section{Discussion}

The present model was intended to gain insight into the evolution of
various social networks by considering mechanisms that account for
heterogeneity in the population of participating entities. To analyse
the statistical properties of the generated network we simplified the
model. We found that the structure of the network dramatically changes
when the number of possible links to a newly added node increases from
\( k=1 \) to \( k=2 \). With \( k=1 \) the network does not form a
giant cluster but the average cluster size goes to infinity (at
$\delta=\frac{1}{2}$) in contrast to \( k\geq 2 \), where the giant
cluster appears in an infinite-order phase transition and the average
cluster size jumps discontinuously but remains finite.  The size of
the jump corresponds to how slowly the giant cluster overcomes the
other competitive large clusters.  However, there is no transition for
\( k=1 \), where none of the clusters can absorb other clusters. The
distribution of the size of finite clusters always follows an
exponential distribution, both below and above the critical point for
$k > 1$, while the model studied in \cite{dorogovtsev01b,callaway01}
is in a critial state below and at the critical point and exhibits an
exponential distribution of cluster size above the transition as in a
Berezinskii-Kosterlitz-Thouless phase transition. Thus, even though
there are disconnected clusters as in our model, there are significant
differences in the behavior of the cluster size distribution.

Our model is similar to a previous model of Callaway
\textit{et al.}  \cite{callaway01}, but there are essential
differences in several points due to nature of the growth 
algorithm: in the model of Callaway \textit{et al.} network
growth and connection formation are independent while in our model only
newly added nodes form connections. Also, in our model multiple
connections might be formed in one time step depending on parameters
$k$ and $\delta$. This difference is well reflected in the generating
function derived for the two models.

The structural properties of our model are more relevant to many social
networks than other growth models such as preferential attachment
because the degree distribution is exponential which is closer to real
social systems and because there are clusters of nodes which represents
the reality of social systems where people usually form various 
communities which are relatively isolated from each other. As long as the
distribution of nodal traits are random, then the structural properties
which we have discussed in this paper do not depend on the nature of
the traits and thus our network model should be relevant to any social
network. The next step is to analyze the distribution of traits on a
social network. This will vary depending on how the attachment rule
depends on the values of these traits even though the structural
properties of the network remains the same. We will discuss the
distribution of traits on a network in a future publication.

\section{Conclusions}

We introduced a model of growing social networks and analyzed its
statistical properties. Our analytical calculations showed that these
growing networks exhibit exponential degree distributions. We gave an
explicit description of the expected number of edges which showed an
exponential dependence on the age of a node. We also showed that
emergence of a giant cluster and the cluster-size distribution
strongly depend on the number of possible initial partners. Numerical
simulations suggested that the generated networks have scale free
cluster distributions only at the phase transition point. In all other
regions of the phase space the cluster distribution was exponential.
In the absence of an exact solution for Eq.~(\ref{eq:gf5}), we showed
numerical results suggesting that the order of the phase transition is
infinite, which is similar to the results found by \cite{callaway01}.

\section*{Acknowledgements}
We acknowledge support from the Henry R. Luce Foundation, the National
Science Research Council (OTKA)  grant No.~T038140, and the National
Science Foundation, grant No.~PHY-9801878.

\end{document}